\documentclass[journal=jacsat,manuscript=communication,layout=traditional]{achemso}
\usepackage[version=3]{mhchem} 

\usepackage{psfrag}
\usepackage{graphicx}
\usepackage{amsmath}
\usepackage{amssymb}
\usepackage{setspace}

\author{Robert Stadler}
\affiliation[TU Wien] {TU Wien - Vienna University of Technology, Institute for Theoretical Physics, Wiedner Hauptstrasse 8-10, A-1040 Vienna, Austria}
\email{robert.stadler@tuwien.ac.at}

\title[\texttt{achemso} Titel]
{Comment on "Breakdown of Interference Rules in Azulene, a Nonalternant Hydrocarbon"}

\begin{document}

Quantum interference (QI) effects in molecular systems are a topic of emerging interest in electron transport studies of single molecule junctions. In a recent Letter, Xia et al.~\cite{xia} employed a graphical scheme introduced by my colleagues and myself~\cite{first}-~\cite{atmol} that enables to distinguish between molecular topologies in conjugated $\pi$ systems which exhibit QI effects in their conductance from those who do not. They claimed that this scheme is not applicable for nonalternant hydrocarbons, in particular Azulenes (Az), whose transport properties were studied both theoretically and experimentally by connecting the electrodes at Az's 1,3-, 2,6-, 4,7-, and 5,7-positions, respectively. An apparent disagreement has been identified between the predictions of our scheme referred to as "atom- or bond-counting model" and theoretical simulations for featureless, wide band electrodes in the GW approximation as well as for the tight-binding (TB) model our method was originally derived from~\cite{first,nano}.

In the following I would like to clarify that this apparent disagreement vanishes if the graphical scheme is applied correctly, providing a different result for the 1,3 Az compound than claimed by Xia et al.~\cite{xia}. It will further be argued that the graphical scheme is completely general for any molecular topology in conjugated $\pi$ systems regardless of whether they are alternant or non-alternant hydrocarbons as long as the basic assumptions of its derivation are fulfilled.

Before addressing the source of error in the reasoning of Ref.~\cite{xia}, I recapitulate the rules we established for complete destructive interference in the wording of Ref.~\cite{nano}: Destructive QI occurs at the Fermi energy if it is impossible to connect the two external atomic sites 1 and N in a molecular topology, i.e. the only two sites with a direct coupling to the electrodes, by a continuous chain of paths, and at the same time fulfill the conditions (i) two sites can be connected by a path if they are nearest neighbours and (ii) at all internal sites, i.e. sites other than 1 and N, there is one incoming and one outgoing path. In other words, for a finite conductance and the absence of QI effects at $E=E_F=0$, all atomic orbitals of the molecular topology have to be either traversed within a continuous chain of paths from 1 to N or be part of a closed loop in the topology, where the latter can be a double line due to the pairing of connected orbitals or a triangle or any larger loop as pointed out explicitly in Refs.~\cite{first} and~\cite{atmol}. 

\begin{figure}
  \includegraphics[width=0.3\columnwidth,angle=0]{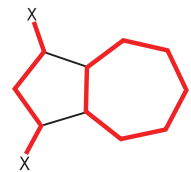}
  \caption[cap.Pt6]{\label{fig1}Graphical scheme from Refs.~\cite{first}-~\cite{atmol} applied to the 1,3-Azulene, where all atomic sites are either part of the continuous path between the connection sites to the electrodes (denoted with X) or part of the closed loop defined by the 7-membered ring and no interference is therefore predicted by our graphical rules. In Ref.~\cite{xia} only the possibility of pairing of atomic sites was checked giving rise to the erroneous claim that the scheme breaks down for this system.
}
  \end{figure}

Xia et al.~\cite{xia} applied these rules correctly for 4,7 Az and 2,6 Az, where continuous paths fulfilling (i) and (ii) could be identified, and also for 5,7 Az where this was not possible. They failed, however, in applying the scheme correctly for 1,3 Az, because they only checked for the possibility of pairing all sites not being part of the main path but did not consider the possibility of a larger closed loop, namely the one provided by the seven membered ring which is illustrated in Fig.~\ref{fig1}. As a result, they concluded wrongly that the graphical interference rules predict QI effects to appear for 1,3 Az and that the rules therefore "break down" for this system. The proper prediction for 1,3 Az from our graphical scheme, however, is that 1,3 Az does not exhibit QI effects in the conductance which is in full agreement with the GW calculations in Fig. 7 of Ref.~\cite{xia} and most crucially also with the TB model calculations in Fig. S4 of the articles supporting informations. For the 5,7 Az compound, which is the only one of all substitution patterns investigated in Ref.~\cite{xia} where our scheme predicts destructive QI if applied correctly, the conductance is indeed found to be exactly zero in Fig. S4 of Ref.~\cite{xia} and reduced by 1-2 orders of magnitude compared to the other compounds as depicted in the articles Fig. 7. It can therefore be concluded that there is no contradiction between the predictions of our graphical scheme and the numerical results in Ref.~\cite{xia}, which removes any foundation for the claim of the authors that our interference rules break down for the non-alternant systems they investigated.

For the case of differently connected Azulene derivatives, our graphical scheme predicts correct QI effects and the proposed interference rules do not break down in contrast of what has been claimed~\cite{xia} as was demonstrated above. It is possible, however, that our graphical rules fail to predict the outcome of conductance experiments if the conditions for its derivations, in particular the requirement of an onsite energy of zero for all atomic sites, are not fulfilled as we addressed in detail in Ref.~\cite{pccp}. Furthermore, they should not be applied in systems where many-body effects play a dominant role requiring a theoretical treatment beyond a single particle level. The graphical method, however, is completely general in predicting QI on the level of the TB model it was derived from and has not yet failed in comparison with density functional theory (DFT) calculations regardless of the molecular topology of the conjugated $\pi$ system. This certainly also applies to non-alternant hydrocarbons and not only to alternant ones, where the differences in the optical excitation spectra between the two types of molecules cannot be translated into predictable trends in single-molecule conductance as the authors concede themselves in Ref.~\cite{xia}. The reason of the general validity of these rules lies in the generality of their derivation which will be briefly recapitulated in the following. 

Formally, any $N\times N$ matrix can be developed according to Leibniz as a sum of permutations,

\begin{equation} \label{eq.leibniz}
\text{det}(A)=\sum_{\sigma}
\text{sgn}(\sigma)\prod_{i=1}^{N}A_{i\sigma(i)}, 
\end{equation}

where $\sigma$ is a permutation of the numbers $1,2,\ldots,N$ and $\text{sgn}(\sigma)$ equals 1 (-1) for an even (odd) permutation. When $A$ equals $H_{\text{mol}}$, i.e. the matrix defining the molecular topology on a TB level in Ref.~\cite{nano}, the condition that all onsite energies of atomic orbitals and therefore all diagonal elements of $A$ are zero has as a consequence that only terms $\prod_{i=1}^{N}A_{i\sigma(i)}$ corresponding to closed loops in the molecular topology remain finite in the summation of Eq.~\ref{eq.leibniz}. When the determinant of $A$ is now alternatively expanded in cofactors according to Laplace, which must result in the same sum of products as in Eq.~\ref{eq.leibniz} albeit not necessarily in the same sequence, it is evident that $a_{1N}$ det$_{1N}(A)$ is the one term in the cofactor expansion, which contains exactly those terms in Eq.~\ref{eq.leibniz} with a direct projection from 1 to N in the respective permutation and therefore invokes the condition of an open continuous path between the two external atomic sites on the products resulting from the determinant of the minor matrix det$_{1N}(A)$, which equals the numerator det$_{1N}(H_{\text{mol}})$ in the expression defining the conductance in Ref.~\cite{nano}. Therefore, the rules (i) and (ii) as named above are completely general for any $N\times N$ matrix $H_{\text{mol}}$ regardless of its topology or size.

In summary, the claim of Xia et al.~\cite{xia} that a graphical scheme my colleagues and I developed for the occurence or absence of QI in the conductance of single molecule junctions with $\pi$ conjugation~\cite{first}-~\cite{atmol} breaks down for non-alternant hydrocarbons such as Azulenes is not justified. They applied the scheme incompletely and therefore incorrectly. Correctly applied, however, the interference rules make correct predictions also for the systems investigated in their article.

{\bf Acknowledgement.} 
R.S. is currently supported by the Austrian Science Fund FWF, project Nr. P27272.

\end{document}